\documentclass[11pt]{article}
\usepackage{graphicx}

\begin{document}

\title{Dynamically Reconfigurable Photonic Crystal Nanobeam Cavities}

\author{Ian W. Frank$^\dagger$, Parag B. Deotare$^\dagger$, Murray W. McCutcheon, Marko Lon\v{c}ar$^*$\\ School of Engineering and
Applied Sciences,\\ Harvard University, Cambridge, MA 02138\\
$^\dagger$  These authors contributed equally
to this work.\\ $^*$ e-mail: loncar@seas.harvard.edu}

\maketitle

\begin{abstract}

Wavelength-scale, high $Q$-factor photonic crystal
cavities~\cite{Song:2005p7806, Deotare:2009p509}
have emerged as a platform of choice for on-chip manipulation of
optical signals, with applications ranging from low-power optical
signal processing~\cite{Tanabe:2005p8072} and
cavity quantum electrodynamics~\cite{Hennessy:2007p8071,
Englund:2007p8179} to biochemical sensing. Many of these
applications, however, are limited by the fabrication tolerances and the
inability to precisely control the resonant wavelength of fabricated
structures. Various techniques for post-fabrication wavelength
trimming \cite{Yang:2007p161114,Faraon:2008p043123} and dynamical
wavelength control -- using, for example, thermal
effects~\cite{Pan:2008p8354, Marki:2006p8348, Fushman:2007p8618}, free
carrier injection~\cite{McCutcheon:2007p10002}, low temperature gas condensation~\cite{Mosor:2005p141105}, and immersion in
fluids~\cite{Maune:2004p589} -- have been explored.
However, these methods are often limited by small tuning ranges,
high power consumption, or the inability to tune continuously or
reversibly. In this letter, by combining nano-electro-mechanical systems (NEMS) and nanophotonics, we demonstrate reconfigurable photonic crystal nanobeam cavities that can be continuously and dynamically tuned using electrostatic forces. A tuning of $\sim 10$ nm has been demonstrated with less than
$6$ V of external bias and negligible steady-state power
consumption.
\end{abstract}

Recently, it has been theoretically
predicted~\cite{Sauvan:2005p8084, McCutcheon:2008p8085,
Notomi:2008p8086} and experimentally
verified~\cite{Deotare:2009p509, Zain:2008p8062,
Eichenfield:2009p4041, Velha:2007p10427} that photonic crystal nanobeam cavities
(PhCNB) can have ultra-high quality factors, on-par with those
demonstrated in conventional photonic crystal cavities based on a
two-dimensional lattice of holes. PhCNB cavities can be viewed as a doubly
clamped nanobeam, the simplest NEMS device, perforated with a
one-dimensional lattice of holes, a textbook example of an optical
grating. By introducing an appropriate chirp in the grating, ultra-high
$Q$ factors and small mode volume optical resonators can be 
realized~\cite{Deotare:2009p509}. When two
PhCNB cavities are placed in each other's near field, as shown in Fig.~\ref{fig:
geometry}, their resonant modes couple, resulting in two supermodes
with resonant frequencies that are highly dependent on the spacing
between the nanobeams~\cite{Deotare09-2}. This can be attributed to two
major factors. First, the coupling between the two resonators
increases with the reduction in the lateral separation between the
nanobeams, which results in greater splitting between the two
supermodes. At the same time, as the nanobeams are drawn closer
together, the higher order effect of the coupling induced frequency
shift~\cite{Popovic:2006p24} becomes significant (especially for
separations $< 100$ nm) and red shifting of both of the supermodes
occurs. The net effect of these two factors is that the even
supermode experiences a considerable red shift as the separation is
reduced, while the wavelength of the odd supermode stays relatively
constant (the two effects cancel out)~\cite{Deotare09-2}.

\begin{figure}[htp]
\centering
\includegraphics[width = .8\linewidth]{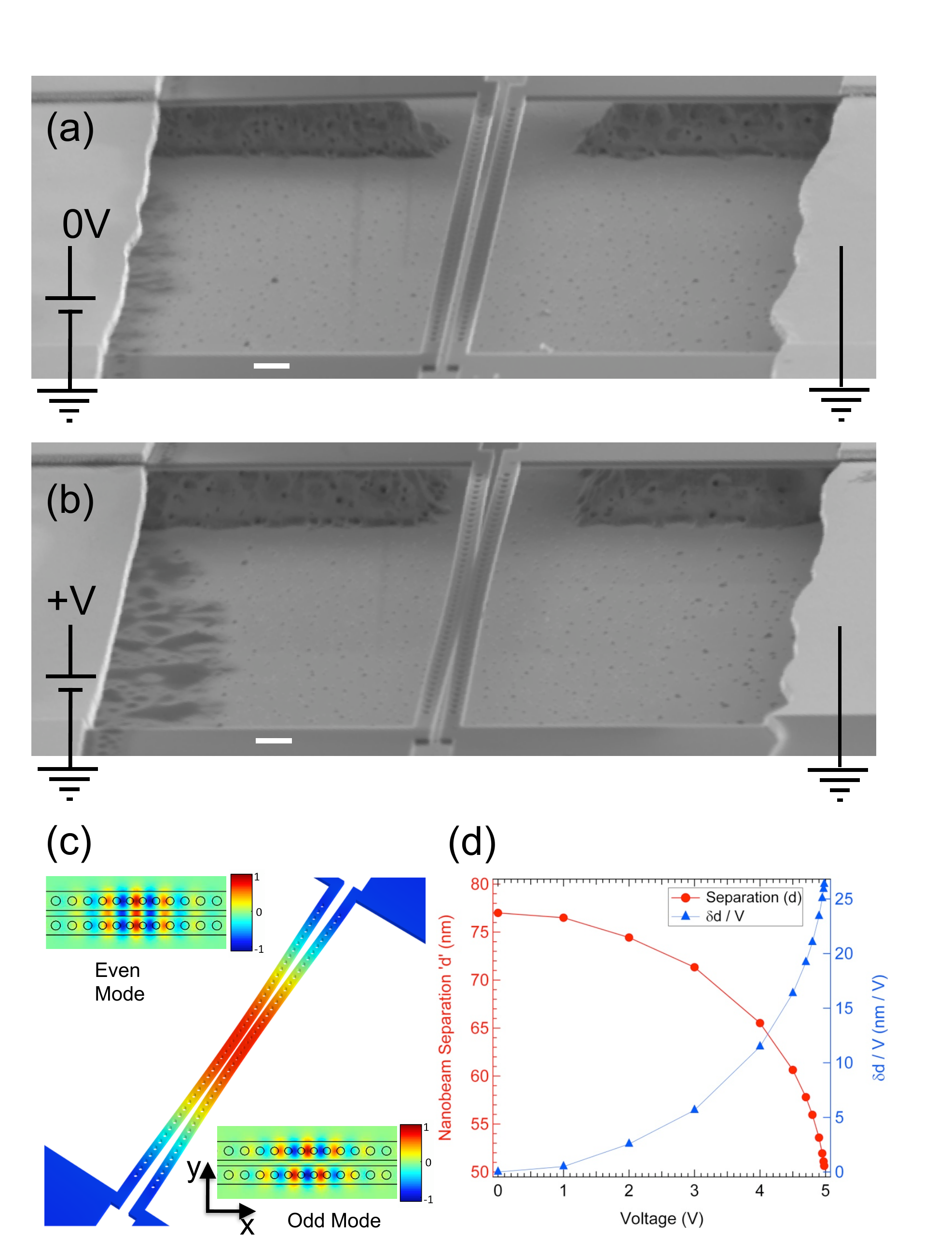}
\caption{\textbf{Coupled photonic crystal nanobeam cavities.} \textbf{a}, SEM image of a representative fabricated structure. The suspended silicon is in contact with gold electrodes seen at the edge of the image and is supported by islands of SiO$_2$ (scalebar = $1~\mu$m). \textbf{b}, SEM image showing the deflection of the nanobeams due to electrostatic actuation. \textbf{c}, Finite element simulations showing nanobeams deflected due to an applied potential. The insets depict the $E_y$ component of the optical supermodes of the coupled cavities. \textbf{d}, Simulation data:  the red curve shows the lateral separation of a pair of nanobeams, measured at the center of the structure, as a potential is applied across them, while the blue curve shows the sensitivity of the deflection with respect to the applied voltage.}
\label{fig: geometry}
\end{figure}

The strong dependence of the wavelength of the even supermode on the separation 
between two nanobeams renders coupled-PhCNB cavities highly suited for 
applications in motion and mass sensing. In addition, the strong optical fields 
that exist in the air region between the coupled-PhCNB cavities makes these devices 
excellent candidates for biochemical sensing applications. Finally, by simultaneously 
taking advantage of both the optical and mechanical degrees of freedom of such 
these cavities, a plethora of exciting optomechanical phenomena can be 
realized~\cite{Eichenfield:2009p4041, Eichenfield:2009arXiv}.

\begin{figure} [hbtp]
\centering
\includegraphics[width = 1\linewidth]{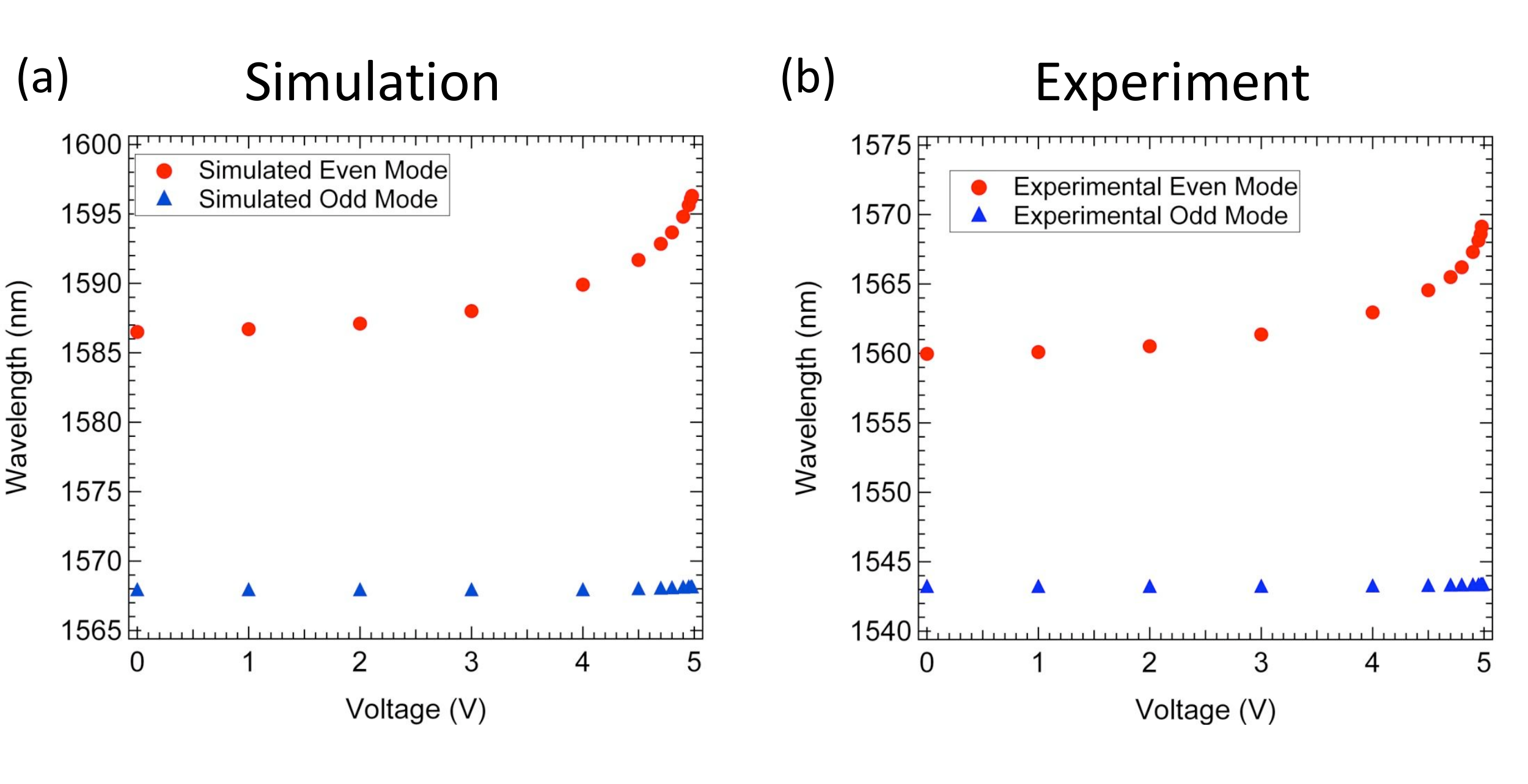}
\caption{\textbf{Electrostatic tuning of a coupled photonic crystal nanobeam cavity.} \textbf{a}, Finite element simulations showing the dependence of the even (shown in red) and odd (blue) supermode resonance on the applied bias voltage. \textbf{b}, Experimental data showing the measured resonances for even and odd supermodes. The trend seen in the experimental data matches well with the simulated results. The slight discrepancy in the absolute value of resonant wavelength can be attributed to uncertainty in the thickness and refractive index of the device layer of the SOI wafer, as well as the amount of tensile stress in the nanobeams.}
\label{fig:tuning}
\end{figure}

In this work, we take advantage of the mechanical flexibility of
coupled PhCNBs to realize reconfigurable optomechanical devices that
can be electrostatically actuated~\cite{Lee:2006p6510}. By applying
a potential difference directly across the nanobeams, an attractive
electrostatic force can be induced between the two nanobeams, resulting in a 
decrease of the gap between the nanobeams, as can be seen in Fig. \ref{fig: geometry} 
b and c. This, in turn, results in the change of the resonant wavelength of the 
two supermodes. Self-consistent optical and mechanical finite-element simulations
were used to model the deflection of the nanobeams due to the electrostatic forces, and 
its influence on the optical eigenfrequencies (Fig.~\ref{fig: geometry}c). 
Fig.~\ref{fig: geometry}d shows the dependence of the nanobeam separation (red curve) 
on the applied voltage, as well as the differential separation change for different
bias voltages, in the case of a device with $77$ nm initial
separation between nanobeams. It can be seen that nanobeam separation, measured at the middle of the nanobeams, can be reduced to 50nm with $\approx5$ V of external bias. Moreover, the change in the separation per unit of applied voltage strongly depends on the applied bias, and is on the order of 25 nm/V for $V\approx5$ V. The influence of the electrostatically-controlled nanobeam separation on the resonances of two supermodes is shown in Fig.~\ref{fig:tuning}a. We found
that, in our system, the even supermode red shifts while the odd
supermode experiences very little dispersion (remaims effectively
stationary). This is in good agreement with our previous
results~\cite{Deotare09-2}, where the dependence of the supermode
eigenfrequencies on lithographically-defined separations (static tuning) was studied.

Encouraged by these results, we fabricated our opto-mechanical devices using similar techniques to those reported in our previous work~\cite{Deotare09-2}. The principal difference here is that the two PhCNBs are electrically isolated. Furthermore, in order to make electrical contact to
each nanobeam, Au contact pads were lithographically patterned onto the
substrate (see Methods section). An electron micrograph of a fabricated structure is shown
in Fig. \ref{fig: geometry}a. The optical characterization of the
fabricated structures was performed using a resonant scattering
setup~\cite{McCutcheon:2005p8090, Altug:2005p982} (see Methods
section). A normally incident tunable laser-beam was focused down
onto the cavities and the resonantly back-scattered signal was analyzed after cross-polarizing
it with respect to the incoming wave. Fig.~\ref{fig:tuning}b shows the experimental results for the
nanobeam cavities, illustrating the dependence of the
even and odd supermode eigenfrequencies on the applied bias voltage.
Very good agreement with numerical modeling can be observed.
The experimentally measured
resonant wavelengths were within $2\%$ of the simulated ones,
and the tuning trend matched very well with the theoretical
predictions. The slight discrepancy can be attributed to several
effects, including the uncertainty in the refractive index of the
doped silicon device layer, variations in the layer thickness, and
uncertainty in the amount of tensile stress in the device layer of
the SOI ($\pm 25$ MPa, according to SOITEC). The optical $Q$ factor 
of the modes was determined by a Fano fit~\cite{Galli:2009p2361}
(see Methods section) to the scattered waveform. The $Q$ factor of the even mode was 
around $13,000$ while that of the odd mode was
around $50,000$. In this work, we intentionally designed and fabricated cavities with a lower $Q$ in order to facilitate experimental characterization via the resonant scattering approach. In our previous work, we demonstrated that coupled PhCNB cavities can have $Q$ factors in the $10^5-10^6$ range~\cite{Deotare09-2}. It is important to emphasize that the $Q$ factors did not change observably across the whole tuning range. This is in stark contrast to tuning via 
free-carrier injection, which results in significant reduction in the cavity $Q$ factor due to
free-carrier absorption.

\begin{figure} [hbtp]
\centering
\includegraphics[width = 1\linewidth]{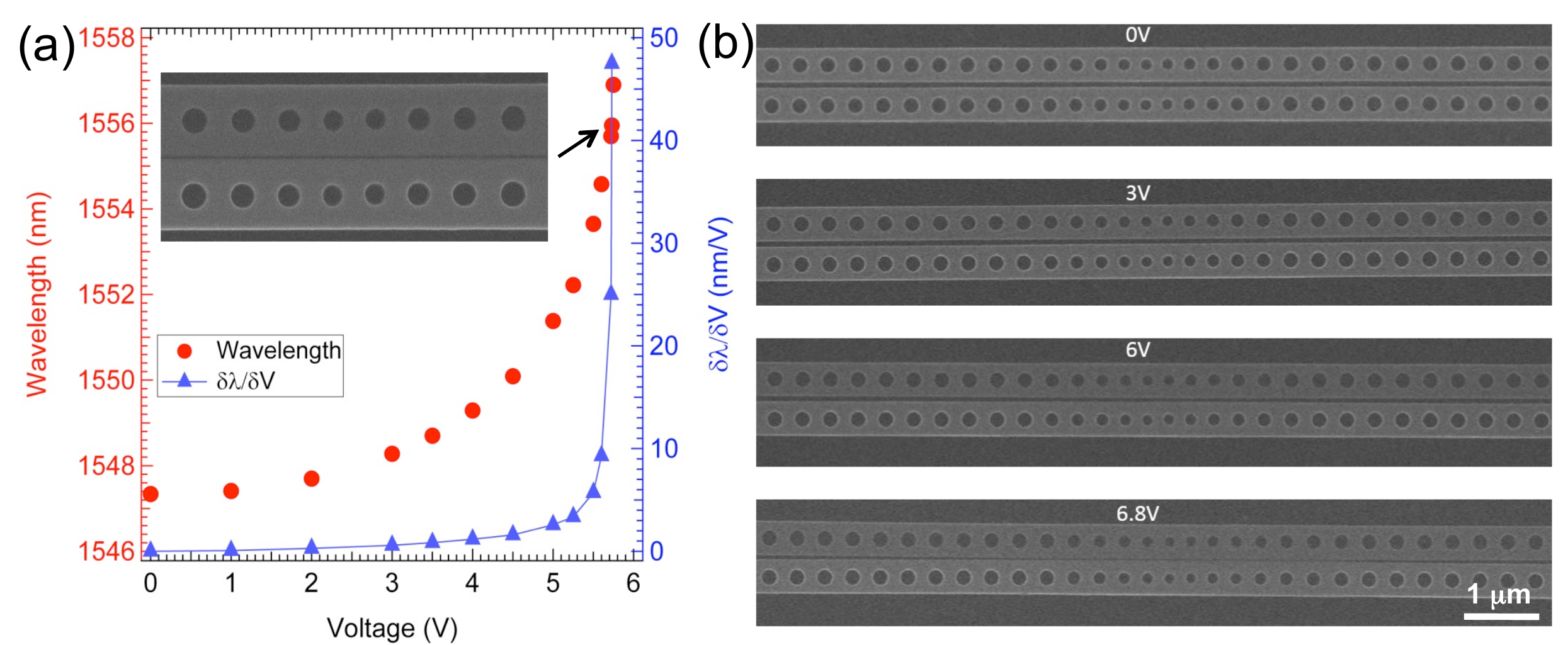}
\caption{\textbf{Sensitivity of the coupled-cavity resonance and visualization of nanobeams deflection due to the applied voltage}. \textbf{a}, Experimental results showing the resonant wavelength of the even supermode as the bias voltage is stepped up to $6$ V (red curve). Tuning up to 9.6nm is obtained in this cavity.
The blue curve shows the sensitivity of the same cavity resonance to the applied voltage. A high sensitivity of $50$ nm/V is obtained when cavity is operated around 6 V bias voltage. The results are obtained for a cavity with initial (V=0) nanobeam separation of $\sim 70$ nm. \textbf{b}, Scanning electron microscopy images showing deflection of a pair of nanobeams under different bias voltages. The lower nanobeam remains grounded, while the potential on the upper nanobeam is increased as indicated.}
\label{fig: semtuning}
\end{figure}

In our best devices, we were able to shift the resonant wavelength of
the even supermode up to $9.6$ nm when less than $6$ V of
external bias voltage was applied (Fig.~\ref{fig: semtuning}a). This wide tuning range
is nearly $80$ times larger than the linewidth of the cavity resonance in the present design, and this ratio can be further improved by increasing the $Q$ factor of the
fabricated cavities. Fig.~\ref{fig: semtuning}a also
shows the sensitivity plot for the measured cavity, defined as the change in the resonant wavelength for a given voltage change. It can be seen that by operating the system at $\sim 6$ V bias voltage, sensitivities as large as $50$ nm/V can be measured. In other words, in this regime, as little as a $5$ mV change in the bias voltage would result in a wavelength change larger than the full-width at half-maximum ($FWHM\sim 0.1$ nm) of the cavity resonance. This is advantageous for the realization of applications such as low-power optical switches and reconfigurable filters/routers. The high sensitivity of our devices can be attributed to two factors: (i) the dependence of the wavelength shift on the change in separation is intrinsically nonlinear~\cite{Deotare09-2}, and much larger shifts are obtained as the nanobeam separation becomes smaller, as in the case of higher voltages; (ii) the electrostatic force experienced by
the nanobeams is quadratic with the applied bias voltage as well as inversely-proportional to the nanobeam separation. It is important to emphasize that in the steady state, when the system is reconfigured and the
nanobeams are deflected to their final position, our system is not
drawing any power from the bias source (beyond what might result
from leakage currents through the thick SiO$_2$ layer underneath the
Si device layer). This is of great practical interest for the realization of reconfigurable devices and systems, as mentioned above. 
The high sensitivities and high $Q$ factors of coupled-PhCNB cavities are also suitable for precision motion measurements in NEMS devices, since a strong modulation of the optical signal can be achieved, even for tiny displacements of the nanobeams. Finally, coupled-PhCNB cavities hold great promise for exciting applications involving the adiabatic
wavelength conversion of light trapped in optical 
cavities~\cite{Preble:2007p8393, Tanabe:2009p8391,McCutcheon:2007p10002} 
through mechanical motion.

By utilizing an electrical feed-through port on a scanning electron microscope (SEM), we were able to observe the real-time deflection of the devices due to the applied bias voltage. Fig.~\ref{fig: semtuning}b shows SEM images of the two nanobeams with increasing voltages applied across them. The images are shown for
nanobeams with a large initial separation ($V_{bias}=0$) of $100$ nm, in order to render the motion of the nanobeams more distinctly. The bending of the nanobeams at the center of the structure can easily be observed, and matches well with our theoretical predictions (Fig.~\ref{fig: geometry}c). After the
pull-in voltage~\cite{liu_foundations_2005} is exceeded, the two
beams can become permanently stuck together due to van der Waals
interactions.  Finally, we note that the difference in steady-state performance of our structures when operated in vacuum (inside the SEM chamber) and in the atmospheric conditions (resonant scattering setup) is negligible, as in either case the structure is operated well below the breakdown voltage.

The current limitation to the tuning method employed here is the RC (resistance $\times$ capacitance) time constant of the system. The resistance offered by the silicon nanobeams is on the order of $10^{11}\,\Omega$, resulting in time
constants in the tens of seconds, despite the small capacitance of
the structure. This could be readily improved by doping the silicon, albeit at the expense of the cavity $Q$ factors (due to increased material loss). The performance of the system could be further improved by utilizing alternative actuation methods~\cite{Unterreithmeier:2009p8790} that do not depend on the RC time constant of the coupled nanobeams. In that case, the response time would be limited by the mechanical response, which is on the order of $\sim 100$ ns in the present design.

In summary, we have demonstrated reconfigurable optical filters that can be dynamically and reversibly tuned using electrostatic forces over $\approx 10$ nm wavelength range when less than 6 V of external bias voltage is applied to the structure. This work will serve as a
basis for exciting applications ranging from ultra-sensitive motion
detection to electrically actuated switching in optical circuits. The
tuning method is stable and remarkably reproducible, provided that the
voltage is not raised beyond the point of pull-in. By allowing precision
wavelength trimming of devices, this method also provides higher tolerances for
fabrication errors, enabling diverse applications in optomechanics, cavity quantum 
electrodynamics, and optical signal processing.

\section{Acknowledgements} This work is supported in part by NSF CAREER grant. Device fabrication was performed
at the Center for Nanoscale Systems at Harvard. The authors would like
to thank CNS Staff members David Lange and Steven Paolini for their
assistance. The authors would also like to thank Prof. Ming Wu for helpful discussions. M.W.M. would like to thank NSERC (Canada) for its support
and I.W.F. thanks the NSF GRFP.

\newpage
\section*{Methods}
\subsection*{Device Fabrication}

The devices were fabricated on a {SOI} substrate with a
$220$ nm device layer using a standard electron beam lithography followed
by an ICP reactive ion etching in a $SF_6$-$C_4F_8$ plasma. The nanobeams were
laterally separated by a distance as small as $50$ nm. Photolithography was used to pattern Cr/Au electrodes for electrical contact with each isolated beam. A thin layer of Cr is used as an adhesive layer for the Au electrodes. A hydrofluoric acid vapor etch was performed to release the structures. Finally, contact was made to the gold electrodes by ultrasonic wirebonding to a ceramic chip-carrier.

\subsection*{Device Characterization}

The devices were characterized using a resonant scattering setup. A {CW} beam was passed through a polarizer and rotated by $45^\circ$ using a half-wave plate before entering the objective lens. The resonantly scattered signal was collected by the same objective lens, split using a non-polarizing beam-splitter, analyzed using a linear polarizer which was cross polarized with respect to the input
beam, and finally detected using an In{G}a{A}s detector. This method enhances the ratio between the resonantly scattered signal and the nonresonant background reflection, without loading the cavity. Due to the inherent symmetry
of the excitation field, the resonant modes of the two cavities are
more naturally driven in phase, which facilitates the measurement of
the even supermodes of the coupled cavities. However, by taking advantage
of a gradient in the excitation fields (by offsetting the excitation beam),
we were also able to probe the odd supermodes.

To obtain the quality factor of the cavity, the scattered signal from the cavity was first normalized
to the background taken away from the cavity but along the beam. The data were then fitted with the Fano lineshape
\begin{equation}
F(\omega) = A_0 + F_0\frac{[q+2(\omega - \omega_0)/\Gamma]^2}{1+[2(\omega - \omega_0)/\Gamma]^2}
\end{equation}
where $\omega$ is the frequency of the cavity mode, $\Gamma$ is the FWHM and $A_0$ and $F_0$ are constants
to obtain the Quality Factor (Q)~\cite{Galli:2009p2361}.

\end{document}